\documentclass[aps,pra,showpacs,twocolumn]{revtex4}
\UseRawInputEncoding 
\usepackage{upgreek}
\usepackage{hyperref}
\usepackage{amsmath}
\usepackage{amssymb}
\usepackage{amsthm}
\usepackage{mathrsfs} 

\usepackage{graphicx}
\usepackage{xcolor}
\newcommand\mb{\boldsymbol}

\begin{document}

\title{On the positive mass theorem in general relativity and Lorentz covariance \\ of the Dirac wave equation in quantum mechanics}
\author{Changbiao Wang }
\email{changbiao\_wang@yahoo.com}
\affiliation{ShangGang Group, 10 Dover Road, North Haven, CT 06473, USA}
%\date{\today}

\begin{abstract}
The positive mass theorem in general relativity states that in an asymptotically flat spacetime, if the momentum--energy tensor is divergence-free and satisfies a dominant energy condition, then a total momentum--energy four-vector can be formed, of which the energy component is nonnegative.  In this paper, we take the wave four-tensor of a plane light wave in free space as a counterexample to show that there is no guarantee that a total four-vector can be formed.  Thus the theoretical framework for the positive mass theorem is flawed. In addition, it is also shown as well that the Lorentz covariance of Dirac wave equation is not compatible with Einstein mass--energy equivalence.
\end{abstract} 

\maketitle 

%Sec. I**************************************
\section{Introduction}
\label{s1}
In the theory of relativity, the momentum and energy of a closed physical system can be described by a four-tensor $T^{\mu\nu}$, usually called momentum--energy tensor.  It is well established in traditional textbooks \cite[p.\,443]{r1} \cite[p.\,46]{r2} \cite[p.\,166]{r3} \cite[p.\,756]{r4} that if the tensor is divergence-free, namely   $T^{\mu\nu}_{~~\,,\nu}=0$, then the total momentum and energy of the system can be obtained by carrying out integration of the time-column elements of the tensor to form a constant four-vector $P^{\mu}=\int T^{\mu4}\textrm{d}^3x$, which is usually called \emph{conservation law}.  However recently, it has been shown by enumerating specific counterexamples that this conclusion is not correct in general \cite{r5,r6,r7,r8}. 

In an asymptotically flat spacetime, if the momentum--energy tensor $T^{\mu\nu}$  satisfies the conservation law $T^{\mu\nu}_{~~\,,\nu}=0$, and also satisfies an additional \emph{dominant energy condition}, namely the energy component $T^{44}\ge |T^{\mu\nu}|$ \cite{r9}, then a total momentum--energy four-vector $P^{\mu}=\int T^{\mu 4}\textrm{d}^3x$  can be formed, of which the total energy component $P^4=\int T^{44}\textrm{d}^3x$  is nonnegative, which is called \emph{positive mass theorem} in the general relativity \cite{r10,r11,r12,r13,r14}.  (Note that Rindler's practice of the tensor indices $\mu, \nu = 1, 2, 3, 4$  \cite[p.\,138]{r15} is used throughout the paper except for in Sec.\,\ref{s2}.)

The best way to disprove a mathematical conjecture is to give its counterexample.  In this paper, we take the wave four-tensor $T^{\mu\nu}=K^{\mu}K^{\nu}$  as a counterexample to show that there is no guarantee that a total four-vector can be formed although it satisfies the traditional conservation law $T^{\mu\nu}_{~~\,,\nu}=0$  and the dominant energy condition  $T^{44}\ge |T^{\mu\nu}|$, where  $K^{\mu}$ is the wave four-vector of a plane light wave in free space, set up by Einstein \cite{r16}.  From this we conclude that the theoretical framework for the positive mass theorem is fundamentally flawed.  

Finally, it is also shown as well that the Lorentz covariance of Dirac wave equation for an electron is not compatible with Einstein mass--energy equivalence. \\

%Sec. II**************************************
\section{Positive mass theorem}
\label{s2}      
In this section, the theoretical framework and conclusions for the positive mass theorem are examined, including the origin of the total energy definition, which is usually omitted in the original research works for various proofs of the theorem \cite{r10,r11,r12,r13,r14}. 

In general relativity, the metric $g_{\mu\nu}$  are the \emph{solutions} of Einstein field equation, while the energy--momentum tensor $T^{\mu\nu}$, which causes space to curve \cite[p.\,5]{r1}, is the \emph{source} of the field equation; just like the EM fields are the solutions of Maxwell equations, while the charge and current are the source of Maxwell equations \cite[p.\,238]{r4}.  (Note that Arnowitt-Deser-Misner practice of the tensor indices $\mu, \nu = 0, 1, 2, 3$  is used in this section for the convenience of readers to check with the source papers \cite{r17,r18}.) 

According to Arnowitt, Deser and Misner \cite{r17}, the definition of the total energy--momentum four-vector $P^{\mu}$  is given by the \emph{volume} integral of the components of  $T^{0\mu}$, namely
\begin{align}
P^{\mu}=\int T^{0\mu}\textrm{d}^3x,
\label{eq1}
\end{align}
which can be expressed as surface integral through Einstein field equations and Gauss theorem \cite[p.\,462]{r1}. $P^0=E$   in above Eq.\,(\ref{eq1}) is the total energy \cite{r18}. 

In the proofs of the positive mass theorem \cite{r11,r12,r14}, the total energy follows the definition in \cite{r17,r18} \cite[p.\,462]{r1}, given by
\begin{align}
E=\int T^{00}\textrm{d}^3x~~=\frac{1}{16\pi}\int (g_{jk,k}-g_{kk,j})\mathrm{d}^2S^j,
\label{eq2}
\end{align} 
where the volume integral is evaluated over the source, and the surface integral is done over a closed surface completely surrounding the source in the asymptotically flat region.  $T^{00}$ satisfies the dominant energy condition $T^{00}\ge |T^{\mu\nu}|$ \cite{r9,r11,r14} and it is of the source term of Einstein field equation, while $(g_{jk,k}-g_{kk,j})$  is the solution related term.

It should be repeatedly emphasized that $\int (...)\,\textrm{d}^3x$  in Eq.\,(\ref{eq1}) means to carry out \emph{volume} integration so that Gauss theorem can be used to convert $\int (...)\,\textrm{d}^3x$ into $\int (...)\,\mathrm{d}^2S^j$ in Eq.\,(\ref{eq2}); confer Ref.\,\cite[p.\,462]{r1}. 

Arnowitt, Deser and Misner claim that because of the conservation law $T^{\mu\nu}_{~~~,\,\mu}=0$, ``$P^{\mu}$ should transform as a four-vector'' \cite{r18}, which is clearly endorsed by Nester \cite{r19}.  In the textbook by Misner, Thorne and Wheeler \cite[p.\,443, p.\,462]{r1}, it is also emphasized that the conservation law  $T^{\mu\nu}_{~~\,,\nu}=0$ ($T^{\mu\nu}_{~~~,\,\mu}=0$) makes $P^{\mu}=\int T^{\mu 0}\textrm{d}^3x$ $~(P^{\mu}=\int T^{0\mu}\textrm{d}^3x)$ be a constant four-vector.

Thus the positive mass theorem actually states that if the tensor $T^{\mu\nu}$  satisfies the conservation law and dominant energy condition, namely $T^{\mu\nu}_{~~~,\,\mu}=0$  and $T^{00}\ge |T^{\mu\nu}|$  hold, then two conclusions can be drawn: Conclusion (a) $P^{\mu}=\int T^{0\mu}\textrm{d}^3x$  is a constant four-vector, and Conclusion (b) $E=P^0=\int T^{00}\textrm{d}^3x$  is nonnegative. 

It is worthwhile to point out that in Eq.\,(\ref{eq2}), no matter in what asymptotically flat region the solution-related surface integral $(1/16\pi)\int (g_{jk,k}-g_{kk,j})\mathrm{d}^2S^j$  is evaluated, it is always equal to the source-related volume integral $\int T^{00}\textrm{d}^3x$; in other words, asymptotically flat solutions  $g_{\mu\nu}$  of Einstein field equation always satisfies Eq.\,(\ref{eq2}).  Thus if $\int T^{00}\textrm{d}^3x$  is not a component of four-vector, then $(1/16\pi)\int (g_{jk,k}-g_{kk,j})\mathrm{d}^2S^j$  is not either.

In the following section, we will prove by enumerating a counterexample that $E=P^0=\int T^{00}\textrm{d}^3x$  is not a component of four-vector, and thus the above Conclusion (a) in the positive mass theorem is not true in general.

%Sec. III**************************************
\section{Counterexample}
\label{s3}   
In this section, a counterexample of the positive mass theorem is provided, which is constructed from the wave four-vector of a plane wave of light in free space.  This wave four-vector, first formulated by Einstein in 1905, predicts the relativistic Doppler effect \cite{r16}, which is the physical basis of the two successive frequency upshifts of a free-electron laser \cite{r20} and has been widely demonstrated by experiments \cite{r21,r22,r23,r24}. 

Suppose that observed in an inertial frame $XYZ$  in free space, the wave four-vector for a plane light wave is given by \cite{r16}
\begin{align}
K^{\mu}=\left(\mathbf{k}_{\textrm{w}}, \frac{\omega}{c}\right),
\label{eq3}
\end{align}
where $K^{1,2,3}=(\mathbf{k}_{\textrm{w}})_{x,y,z}$,  $K^4=\omega/c$, $\mathbf{k}_{\textrm{w}}$  is the wave vector,  $\omega ~(>0)$   is the angular frequency, and $c$ is the speed of light in free space.

The wave four-vector satisfies  $K^{\mu}K_{\mu}=0$, and we have  $(\omega/c)^2-\mathbf{k}_{\textrm{w}}^2=0\Rightarrow |\mathbf{k}_{\textrm{w}}|=\omega/c=K^4$.  With $K^4=|\mathbf{k}_{\textrm{w}}|=[(\mathbf{k}_{\textrm{w}})_x^2+(\mathbf{k}_{\textrm{w}})_y^2+(\mathbf{k}_{\textrm{w}})_z^2]^{1/2}$  and    $(\mathbf{k}_{\textrm{w}})_{x,y,z}=K^{1,2,3}$ taken into account, we have
\begin{equation}
K^4\ge |K^{1,2,3}|.
\label{eq4}
\end{equation}

As a counterexample of the positive mass theorem, the wave four-tensor is defined as
\begin{equation}
T^{\mu\nu}=K^{\mu}K^{\nu}~\textrm{for}~\mathbf{x}\in V;\quad = 0~\textrm{for}~\mathbf{x}\not\in V;
\label{eq5}
\end{equation}
where the finite volume $V$ $(\neq 0)$  is fixed in  $XYZ$.  According to above Eq.\,(\ref{eq5}), the wave four-tensor $T^{\mu\nu}$  is a finite distribution source of Einstein field equation. Because $K^{\mu}$  is independent of space and time variables $X^{\mu}=(\mathbf{x}, ct)$, $T^{\mu\nu}$ is divergence-free within the volume  $V$, namely 
\begin{equation}
T^{\mu\nu}_{~~\,,\nu}=\frac{\partial T^{\mu\nu}}{\partial X^{\nu}}=\frac{\partial (K^{\mu}K^{\nu})}{\partial X^{\nu}}=0.\\[5pt]
\label{eq6}
\end{equation} 

From Eqs.\,(\ref{eq4}) and (\ref{eq5}) we find that $T^{\mu\nu}$  satisfies the dominant energy condition in $V$, 
\begin{equation}
T^{44}=(K^4)^2\ge |K^{\mu}K^{\nu}|=|T^{\mu\nu}|.
\label{eq7}
\end{equation} 

Consider the volume integral of the components of $T^{\mu 4}$ over $V$, given by 
\begin{align}
P^{\mu}&=\int_V T^{\mu 4}\textrm{d}^3x
\notag \\[2pt]
&=\int_V (K^{\mu}K^4)\textrm{d}^3x=(K^4V)K^{\mu},
\label{eq8}
\end{align} 
where $\int_V (K^{\mu}K^4)\,\textrm{d}^3x=(K^4V)K^{\mu}$  is employed because $K^{\mu}$  is independent of $X^{\mu}=(\mathbf{x},ct)$.

From above Eqs.\,(\ref{eq6}) and (\ref{eq7}), we know that the wave four-tensor $T^{\mu\nu}$  satisfies the conservation law $T^{\mu\nu}_{~~\,,\nu}=0$  and the dominant energy condition $T^{44}\ge|T^{\mu\nu}|$ over $V$.  Thus according to the positive mass theorem, the quantity $P^{\mu}=(K^4V)K^{\mu}$, given by Eq.\,(\ref{eq8}), must be a Lorentz covariant four-vector. 

Thus if we can prove that $P^{\mu}=(K^4V)K^{\mu}$ is not a Lorentz four-vector, then $T^{\mu\nu}$ becomes a counterexample of the positive mass theorem.  As shown below,  $P^{\mu}=(K^4V)K^{\mu}$ indeed cannot be a four-vector. 

Suppose that the inertial frame $X'Y'Z'$  moves with respect to $XYZ$  at an arbitrary velocity of $\mathbf{v}=\mb{\beta}c$, with $X'Y'Z'$  and  $XYZ$  overlapping at $t'=t=0$ \cite{r6}.  Observed in  $X'Y'Z'$, the volume integral of the components of $T'^{\mu 4}$  is given by
\begin{align}
P'^{\mu}&=\int_{V'} T'^{\mu 4}\textrm{d}^3x'
\notag \\[2pt]
&=\int_{V'} (K'^{\mu}K'^4)\textrm{d}^3x'=(K'^4V')K'^{\mu},
\label{eq9}
\end{align} 
where the volume $V'$  is moving with respect to $X'Y'Z'$  at the velocity $\mathbf{v}'=\mb{\beta}'c=-\mb{\beta}c$.  

Comparing above Eq.\,(\ref{eq9}) with Eq.\,(\ref{eq8}), we find that if the relation between $P^{\mu}=(K^4V)K^{\mu}$  and $P'^{\mu}=(K'^4V')K'^{\mu}$  follows four-vector Lorentz transformation, then   $(K^4V)=(K'^4V')$ = \emph{Lorentz covariant scalar} must hold, because $K^{\mu}$ is a four-vector.  However as shown below, $(K^4V)=(K'^4V')$ cannot hold for an \emph{arbitrary}  $\mathbf{v}=\mb{\beta}c$, namely $(K^4V)$  is not a covariant scalar, and so $P^{\mu}=(K^4V)K^{\mu}$  cannot be a four-vector.

According to the \emph{change of variables theorem} in principles of classical mathematical analysis \cite[p.\,252]{r25}, the transformation of differential elements appearing in Eq.\,(\ref{eq8}) and Eq.\,(\ref{eq9}) is given by  
\begin{equation}
\textrm{d}^3x'=\left|\frac{\partial(x',y'z')}{\partial(x,y,z)}\right|\textrm{d}^3x=
\frac{1}{\gamma}\textrm{d}^3x,
\label{eq10}
\end{equation} 
where $\gamma=(1-\mb{\beta}^2)^{-1/2}$ is the Lorentz factor, and the Jacobi determinant $\partial(x',y'z')/\partial(x,y,z)=1/\gamma$  is employed.  It should be noted that the transformation of above Eq.\,(\ref{eq10})  meets the physical requirement of Lorentz contraction effect of the length of a moving rigid rod argued by Einstein \cite{r16} and further formulated in Ref.\,\cite[Footnote 2 there]{r26}. 

From above Eq.\,(\ref{eq10}), we have 
\begin{equation}
\int_{V'}\textrm{d}^3x'=\int_V\left(\frac{1}{\gamma}\right)\textrm{d}^3x \quad\quad \textrm{or} \quad\quad  V'=\frac{V}{\gamma}.
\label{eq11}
\end{equation} 

According to Einstein \cite{r16}, the Lorentz transformation between $K'^4=\omega'/c$  and  $K^4=\omega/c$ is given by 
\begin{equation}
K'^4=\gamma(K^4-\mathbf{k}_{\textrm{w}}\cdot\mb{\beta}).
\label{eq12}
\end{equation} 

Combining Eqs.\,(\ref{eq11}) and (\ref{eq12}), we have 
\begin{equation}
K'^4V'=K^4V-(\mathbf{k}_{\textrm{w}}\cdot\mb{\beta})V,
\label{eq13}
\end{equation} 
which is valid for any $|\mb{\beta}|<1$.

Obviously, $(\mathbf{k}_{\textrm{w}}\cdot\mb{\beta})V=0$  cannot hold for arbitrary $|\mb{\beta}|<1$; for example, $(\mathbf{k}_{\textrm{w}}\cdot\mb{\beta})V=(\omega/c)|\mb{\beta}|V \ne 0$  holds for $\mathbf{k}_{\textrm{w}}\|\,\mb{\beta}$ and  $\mb{\beta}\ne 0$.  Thus from  Eq.\,(\ref{eq13}) we conclude that $(K^4V)=(K'^4V')$  cannot hold for arbitrary  $|\mb{\beta}|<1$, namely $(K^4V)$  is not a Lorentz covariant scalar.  Since $(K^4V)$  is not a covariant scalar, the quantity $P^{\mu}=(K^4V)K^{\mu}$, given by Eq.\,(\ref{eq8}), cannot be a four-vector.   

So far, we have finished the proof that the wave four-tensor $T^{\mu\nu}$ satisfies $T^{\mu\nu}_{~~\,,\nu}=0$  and  $T^{44}\ge|T^{\mu\nu}|$ over the volume $V$, but $P^{\mu}=\int_V T^{\mu 4}\textrm{d}^3x$  is not a four-vector ($\Leftrightarrow$ $P^{4}=\int_V T^{44}\textrm{d}^3x$ is not a component of four-vector); namely the wave four-tensor $T^{\mu\nu}$, defined in Eq.\,(\ref{eq5}), is a counterexample of the positive mass theorem.

%Sec. IV**************************************
\section{Conclusions and remarks}
\label{s4}  
By examining the background, we have found that the positive mass theorem has: 
\begin{itemize}
\item  \emph{two requirements}: the momentum--energy tensor $T^{\mu\nu}$, as the source of Einstein field equation, satisfies (a) the conservation law  $T^{\mu\nu}_{~~\,,\nu}=0$ and (b) the dominant energy condition  $T^{44}\ge|T^{\mu\nu}|$; 
\item \emph{two conclusions}: (a) the time-column volume integral $P^{\mu}=\int T^{\mu 4}\textrm{d}^3x$  constitutes a total four-momentum that is conserved \cite[p.\,443]{r1}, and (b) the total energy $E=P^4=\int T^{44}\textrm{d}^3x$  is nonnegative.  
\end{itemize}

We have proved that the wave four-tensor, given in Eq.\,(\ref{eq5}), is a counterexample of the positive mass theorem because it satisfies the above two requirements, but Conclusion (a) does not hold.  Since Conclusion (a) is not valid, Conclusion (b) naturally loses its physical significance.  Thus the positive mass theorem is fundamentally flawed.

From the counterexample provided in the present paper, one can see that the problem of positive mass theorem comes from the traditional conservation law  $T^{\mu\nu}_{~~\,,\nu}=0$, which is actually an incorrect conjecture \cite{r5,r6,r7,r8} although some proofs in support of it are provided in textbooks, such as in \cite[p.\,142]{r1} \cite[p.\,318]{r27}. 

In addition to the conservation law $T^{\mu\nu}_{~~\,,\nu}=0$, there are some other basic concepts and definitions which have not yet reached a consensus, such as the covariance of physical laws \cite{r28}, electromagnetic power flow in an anisotropic medium \cite{r29}, and Fermat principle for a plane light wave \cite{r30}.  Probably one of the best known and most controversial examples is about how to define a particle's mass.  In respected textbooks \cite{r15,r31}, two different mass definitions are argued, which leads to an amazing and unexpected result: the Lorentz covariance of Dirac wave equation is not compatible with Einstein mass-energy equivalence, as shown below.  

In the first mass definition for a particle, the four-momentum is given by \cite[p.\,110]{r15} 
\begin{align}
P^{\mu}=\frac{m}{\gamma}U^{\mu}&=\left(m\mathbf{v},~\frac{mc^2}{c}\right)
\notag \\[3pt]
&=\left(\mathbf{p},~\frac{E}{c}\right), \quad (\textrm{First definition})
\label{eq14}
\end{align} 
where $m$  is defined as the particle's (relativistic inertial) mass; $(m/\gamma)$  is a \emph{covariant} scalar (invariant) \cite{r28}; $U^{\mu}=\gamma(\mathbf{v},c)$  is the four-velocity, with $\mathbf{v}$ the particle's velocity; $\mathbf{p}$  is the momentum; and $E$  is the energy.  In this definition, (i) $(m/\gamma)=m_0$  holds, where $m_0$ is the rest mass and it is $not$ a covariant invariant \cite{r28}; (ii) Einstein mass--energy equivalence equation $E=mc^2$ holds in all inertial frames.  

In the second mass definition, the particle's four-momentum is given by \cite[p.\,289]{r31} 
\begin{align}
P^{\mu}=mU^{\mu}&=\left(\gamma m\mathbf{v},~\frac{\gamma mc^2}{c}\right)
\notag \\[3pt]
&=\left(\mathbf{p},~\frac{E}{c}\right), \quad (\textrm{Second definition})
\label{eq15}
\end{align} 
where $m$  is defined as the particle's (invariant) mass and it is a covariant invariant; $\mathbf{p}=\gamma m\mathbf{v}$  is the momentum; and $E=\gamma(mc^2)$  is the energy.  In this definition, (i) $m=m_0$  always holds in all inertial frames (namely $m_0$  and $m$  are the same), and $m_0$  is a covariant invariant; (ii) $E=\gamma(mc^2)$  holds, but the Einstein mass--energy equivalence equation $E=(mc^2)$ does not hold except for in the particle-rest frame.

According to the first mass definition shown in Eq.\,(\ref{eq14}), the particle's rest mass is not a covariant invariant; thus Dirac wave equation for an electron is not covariant because the electron's rest mass appearing in Dirac equation is not a covariant invariant although Dirac took it to be in his proof \cite[p.\,258]{r32}.  However the covariance of Dirac equation is consistent with the second definition shown in Eq.\,(\ref{eq15}), which contradicts with Einstein mass--energy equivalence equation.  From this one can conclude that the Lorentz covariance of Dirac wave equation is not compatible with Einstein mass--energy equivalence.  In other words, if Einstein mass--energy equivalence is valid, then Dirac wave equation is not Lorentz covariant, and vice versa.

It is worthwhile to point out that in the first mass definition, the photon has its (kinetic) mass \cite{r28} because ``the mass of a body is a measure of its energy content'' \cite{r33}, while in the second mass definition, the photon does not have any mass (equal to zero) \cite[p.\,99]{r34}.  However Einstein's thought experiment for mass--energy equivalence does support the view that the photon has a non-zero mass, as shown below.

In his thought experiment \cite{r33}, Einstein assumes that a stationary body with a rest energy of $E_0$   emits two identical photons (``plane waves of light'' in Einstein's words) in free space at the same time in the opposite directions so that the body keeps at rest after the emissions.  According to the energy conservation law, Einstein argues  
\begin{equation}
E_0=(E_0-\Delta E_0)+\frac{1}{2}\Delta E_0+\frac{1}{2}\Delta E_0,
\label{eq16}
\end{equation} 
where $(E_0-\Delta E_0)$ is the body's energy after emissions, and the last two terms of $(\Delta E_0/2)$   are the (kinetic) energies of the two photons respectively.  Dividing above Eq.\,(\ref{eq16}) by $c^2$  yields 
\begin{equation}
\frac{E_0}{c^2}=\left(\frac{E_0}{c^2}-\frac{\Delta E_0}{c^2}\right)+\frac{1}{2}\frac{\Delta E_0}{c^2}+\frac{1}{2}\frac{\Delta E_0}{c^2}.
\label{eq17}
\end{equation} 

In terms of the principle of relativity, Einstein proved that the body's rest mass is reduced by $(\Delta E_0/c^2)$  after emissions.  Thus it is legitimate to define $(E_0/c^2)$  and $(E_0-\Delta E_0)/c^2$  as the masses of the stationary body before and after the emissions.  From this it follows that the last two terms of  $\Delta E_0/(2c^2)$ in Eq.\,(\ref{eq17}) legitimately denote the masses of the two photons respectively. 
 
Thus Einstein's thought experiment \cite{r33} supports the conclusion that the photon has a non-zero kinetic energy, and it has a non-zero mass; namely the mass and energy are equivalent, as Einstein argued.

\end{document}